\documentclass[aps,pre,twocolumn,showpacs,superscriptaddress]{revtex4}

\usepackage{graphicx,subfigure}
\usepackage{amssymb}
\usepackage{amsmath}
\usepackage[british]{babel}

\begin{document}

\title{Non-ergodicity, fluctuations, and criticality in heterogeneous diffusion
processes}

\author{A. G. Cherstvy}
\affiliation{Institute for Physics \& Astronomy, University of Potsdam,
14476 Potsdam-Golm, Germany}
\author{R. Metzler}
\affiliation{Institute for Physics \& Astronomy, University of Potsdam,
14476 Potsdam-Golm, Germany}
\affiliation{Department of Physics, Tampere University of Technology, 33101
Tampere, Finland}

\date{\today}

\begin{abstract}
We study the stochastic behavior of heterogeneous diffusion processes with the
power-law dependence $D(x)\sim|x|^{\alpha}$ of the generalized diffusion
coefficient encompassing sub- and superdiffusive anomalous diffusion. Based on
statistical measures such as the amplitude scatter of the time averaged mean
squared displacement of individual realizations, the ergodicity breaking and
non-Gaussianity parameters, as well as the probability density function $P(x,t)$
we analyze the weakly non-ergodic character of the heterogeneous diffusion process
and, particularly, the degree of irreproducibility of individual realization. As
we show, the fluctuations between individual realizations increase with growing
modulus $|\alpha|$ of the scaling exponent. The fluctuations appear to diverge
when the critical value $\alpha=2$ is approached, while for even larger $\alpha$
the fluctuations decrease, again. At criticality, the power-law behavior
of the mean squared displacement changes to an exponentially fast growth, and
the fluctuations of the time averaged mean squared displacement do not seem to
converge for increasing number of realizations. From a systematic comparison
we observe some striking similarities of the heterogeneous diffusion process with
the familiar subdiffusive continuous time random walk process with power-law
waiting time distribution and diverging characteristic waiting time.
\end{abstract}


\maketitle

\section{Introduction}

Over the last fifteen years there has been a surge of studies in anomalous
diffusion, characterized by the deviation of the mean squared displacement
(MSD)
\begin{equation}
\langle x^2(t)\rangle=\int_{-\infty}^{\infty}x^2P(x,t)dx
\end{equation}
of a stochastic process with the probability density function $P(x,t)$ to find
the particle at position $x$ at time $t$, from the linear time dependence $\langle
x^2(t)\rangle=2Dt$ of ordinary Brownian motion \cite{vankampen}. Anomalous
diffusion is usually characterized in terms of the power-law form
\begin{equation}
\label{msd}
\langle x^2(t)\rangle\sim2D_{\beta}t^{\beta}
\end{equation}
with the anomalous diffusion coefficient $D_{\beta}$ of physical dimension
$\mathrm{cm}^2/\mathrm{sec}^{\beta}$ and the anomalous diffusion exponent
$\beta$. Depending on the value of $\beta$ we distinguish subdiffusion
($0<\beta<1$) and superdiffusion ($\beta>1$) \cite{report,bouchaud}.

Since the milestone discoveries of superdiffusion in turbulence already in 1926
\cite{richardson} and of subdiffusion in amorphous semiconductors \cite{scher}
the recent vast increase of interest in anomalous diffusion is due to its
discovery in numerous microscopic systems, in particular in biological contexts.
The cytoplasm of biological cells is heavily crowded with various obstacles,
including proteins, nucleic acids, ribosomes, the cytoskeleton, as well as internal
membranes compartmentalizing the cell \cite{crowd,crowd2}. Diffusion of natural
and artificial tracers in this complex environment is often subdiffusive. Similar
situations are encountered in cell membranes \cite{zhou,weiss}. The experimental
evidence for subdiffusion in the crowded cytoplasm of living cells ranges from the
motion of small labeled proteins \cite{lang-subdiff-nucleus-fcs,fradin05} over mRNA
molecules and chromosomal loci \cite{goldingcox,weber}, lipid and insulin granules
\cite{granules,insulin}, virus particles \cite{brauch01,brauch02}, as well as
chromosomal telomeres \cite{telosubdiff} and Cajal bodies \cite{platani} inside
the nucleus. Subdiffusion was observed for membrane resident proteins
experimentally \cite{weiss,channels} and for membrane lipid molecules in
computer simulations \cite{ad-lipids}. In controlled \emph{in vitro\/}
experiments with artificial crowders, anomalous diffusion was consistently
observed \cite{szymanski,jae_njp}. On larger scales anomalous diffusion was
observed, for instance, for the motion of bacteria in a biofilm \cite{biofilm}.

The observed subdiffusion was ascribed to various physical mechanisms
\cite{saxton,sokolov12-sm,weiss-AD-SM,pt,franosch13}. Apart from the apparent
transient anomalous diffusion caused by a crossover from free normal diffusion
to the plateau value of the MSD \cite{saxton,berez}, typically, three main families
of anomalous diffusion processes are considered: (i) diffusion in a fractal
environment where dead ends and bottlenecks slow down the motion on all scales
\cite{fractal}; (ii) motion in a viscoelastic environment, in which the effective
anomalous motion of the tracer particle in the correlated many-body environment
shows long-ranged antipersistent motion \cite{goychuk}. The latter process is
associated with fractional Brownian motion (FBM) and generalized Langevin equation
motion with a power-law memory form of the friction kernel \cite{fbm,fle}. (iii)
And continuous time random walk (CTRW) models, in which the moving particle is
successively trapped by binding events to the environment or caging effects for
waiting times $\tau$ distributed like a power-law $\psi(\tau)\simeq\tau^{-1-\beta}$
with $0<\beta<1$ \cite{scher,montroll}. All three mechanisms lead to the power-law
MSD (\ref{msd}) and they were indeed identified as processes generating the motion
of different tracers in different cellular environments \cite{weber,granules,
insulin,channels,ad-lipids,szymanski,jae_njp,sokolov12-sm,weiss-AD-SM,pt,
franosch13} or in colloidal systems \emph{in vitro\/} \cite{xu}.

\begin{figure}
\includegraphics[width=8cm]{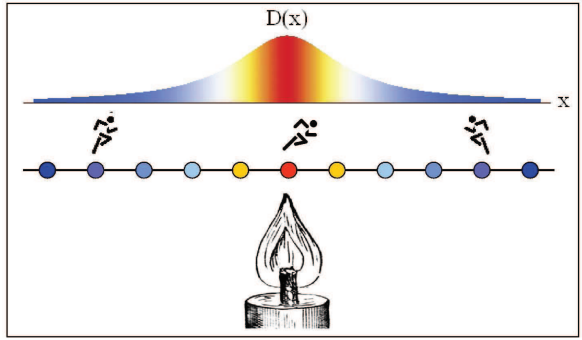}
\caption{Schematic for the spatially varying diffusivity due to temperature or
mobility gradients, shown here for the case of subdiffusion with $D(x)=D_0/
(x_{\mathrm{off}}^2+x^2)$. The diffusivity of the random walker is coupled to
the spatial variation of the
temperature of the environment, as symbolized by the speed of the runner.} 
\label{fig-model}
\end{figure}

These three anomalous diffusion mechanisms characterized by anomalous diffusion
with a constant in time generalized diffusivity were identified in experiments
involving fairly large endogenous as well as artificial tracers. Recent experiments
on
eukaryotic cells \cite{lang11} using considerably smaller tracer proteins sampling
over much larger subvolumes of the cell indicated a systematic variation of the
cytoplasmic diffusivity with the separation from the cell nucleus. These spatial
diffusivity gradients are due partly to the non-uniform distribution of crowders in
the cytoplasm. This distribution can non-trivially affect the diffusion of tracers
of different sizes in the cell cytoplasm. \emph{In vitro}, fast gradients of the
diffusivity can be realized, for instance, via a local variation of the temperature
in thermophoresis experiments \cite{therm1,therm2}, as sketched in
Fig.~\ref{fig-model}. The
diffusion of Brownian particles in explicit solvents with temperature gradients
was recently studied by multi-particle collision dynamics \cite{ripoll-12}.
Fluctuation-dissipation relations and spurious drift effects in systems with
spatially varying friction coefficient were recently treated theoretically in
Ref.~\cite{farago14-friction}. On larger scales, the diffusion of water molecules
monitored by diffusive Magnetic Resonance Imaging in the brain white matter was
demonstrated to be heterogeneous and strongly anisotropic \cite{brain-diffusion}.
The anisotropy is due to the presence of some spatially-oriented structures in
the tissue and obstacles which give rise to a tensorial character of the apparent
diffusion coefficient. The existence of a population splitting into two pools of
water molecules with slow and fast diffusivities was shown for brain white matter
\cite{brain-diffusion}. Finally, spatial heterogeneities are also abundant in the
completely different context of anomalous diffusion in subsurface hydrology
\cite{hydro}.

In the present study we examine the effects of the strength of the diffusivity
gradient as defined by the scaling exponent $\alpha$ and of the initial particle
position $x_0$ in heterogeneous diffusion processes (HDPs) with space-dependent
diffusion coefficient $D(x)$. We pay particular attention to a phenomenon, that
recently received considerable attention for its immediate relevance to the surging
field of single particle tracking experiments in microscopic systems, namely, the
so-called \emph{weak ergodicity breaking}. This is the distinct disparity between
physical observables depending on whether they are evaluated in the conventional
ensemble sense or, from measured time series $x(t)$ of the particle position, in
terms of time averages \cite{russians-eb,bouchaud-eb,pt,sokolov12-sm,franosch13}.
We find that the HDP process gives rise to weakly non-ergodic behavior with a
pronounced amplitude scatter of the time averaged MSD of individual trajectories.
We obtain details of the distribution of this scatter as well as the frequently
used ergodicity breaking and non-Gaussianity parameters. Our analysis uncovers
remarkable similarities of the HDP process with those of CTRW motion. In particular,
we study the behavior of the HDP at the critical value of the scaling exponent of
the diffusivity.

The paper is organized as follows. We introduce the HDP model with $x$-dependent
diffusivity in Sec.~\ref{sec-model}. The main results for the evolution of the MSD,
the time averaged MSD, the probability density function, the ergodicity breaking
and non-Gaussianity parameters in the whole range of the model parameters are then
presented in Sec.~\ref{sec-main-results}. We discuss our results and point out the
directions for future research in Sec.~\ref{sec-discussion}.

\section{Heterogeneous diffusion processes}
\label{sec-model}

HDPs are defined in terms of the multiplicative yet Markovian Langevin equation
\cite{hdp}
\begin{equation}
\label{lang}
\frac{d}{dt}x(t)=\sqrt{2D(x)}\zeta(t),
\end{equation}
where $D(x)$ is the position-dependent diffusion coefficient and $\zeta(t)$
represents white Gaussian noise. In what follows we concentrate on the power-law
form
\begin{equation}
D(x)\sim D_0(|x|^{\alpha}+|x_{\mathrm{off}}|^{\alpha})\sim D_0|x|^{\alpha}
\label{eq-Dx}
\end{equation}
for the diffusivity, where the amplitude $D_0$ has dimension $\mathrm{cm}^{2-
\alpha}/\mathrm{sec}$. Logarithmic and exponential forms for $D(x)$ were
considered in Ref.~\cite{hdp}. The offset $x_{\mathrm{off}}$ in Eq.~(\ref{eq-Dx})
avoids either divergencies of $D(x)$ ($\alpha<0$) or stalling of the particle
($\alpha>0$) around $x=0$ in the simulations. In the following calculations we
use the scaling form $D(x)\sim D_0|x|^{\alpha}$. We interpret the Langevin equation
(\ref{lang}) in the Stratonovich sense \cite{vankampen,hdp}.

The MSD following from the stochastic equation (\ref{lang}) with diffusivity
(\ref{eq-Dx}) takes on the power-law form \cite{hdp} 
\begin{equation}
\left<x^2(t)\right>=\frac{\Gamma(p+1/2)}{\sqrt{\pi}}\left(\frac{2}{p}\right)^{2p}
(D_0t)^p
\label{eq-msd-scaling}
\end{equation}
where we introduced the scaling exponent
\begin{equation}
\label{pexp}
p=\frac{2}{2-\alpha},
\end{equation}
which denotes superdiffusion for $2>\alpha>0$ and subdiffusion for $\alpha<0$. For
$\alpha>0$ the diffusivity grows away from the origin, leading to a progressive
acceleration of the particle as it ventures into more distant regions from the
origin, and vice versa for $\alpha<0$. In the special case $\alpha=2$, the theory
developed in Ref.~\cite{hdp} breaks down as the scaling exponent (\ref{pexp})
diverges. In Refs.~\cite{fulinski,lubensky07} an exponential growth of the MSD was
found, see the discussion below. For even larger values of $\alpha$ the anomalous
diffusion exponent $p$ becomes negative, i.e., we observe a strong localization,
see the discussion below. The probability density function (PDF) of the HDP
is given by the stretched or compressed Gaussian \cite{hdp}
\begin{equation}
\label{PDF-HDP}
P(x,t)=\frac{|x|^{-\alpha/2}}{\sqrt{4\pi D_0t}}\exp\left(-\frac{|x|^{2-\alpha}}{
(2-\alpha)^{2}D_0t}\right).
\end{equation}
For positive $\alpha$ it exhibits a cusp, while for negative $\alpha$ the PDF
features a dip to zero at the origin.

In single particle tracking experiments one measures the time series $x(t)$ of the
particle position for a time span $T$. It is usually evaluated in terms of the time
averaged MSD \cite{pt,sokolov12-sm,franosch13}
\begin{equation}
\overline{\delta^2(\Delta)}=\frac{1}{T-\Delta}\int_0^{T-\Delta}\Big[x(t+\Delta)-
x(t)\Big]^2dt,
\label{eq-tamsd}
\end{equation}
where $\Delta$ is the lag time. Brownian motion is ergodic and for sufficiently
long measurement times $T$ the equality $\langle x^2(\Delta)\rangle=\overline{
\delta^2(\Delta})$ holds \cite{pccp,he}. Anomalous diffusion (\ref{msd}) described
by FBM or fractional Langevin equation motion is asymptotically ergodic
\cite{deng,goychuk} but may feature transiently non-ergodic behavior
\cite{jae_pre,jae_njp} in confinement as well as transient aging \cite{jochen},
the explicit dependence on the time span elapsing between initial preparation of
the system and start of the recording of the position, $x(t)$.

CTRW processes with diverging time scales of the waiting time distribution $\psi(
\tau)$ show weakly non-ergodic behavior for all $\Delta$. Namely, despite the
scaling (\ref{msd}) of the MSD, $\overline{\delta^2}$ scales \emph{linearly\/} with
$\Delta$. More precisely, if we average over sufficiently many trajectories, the
quantity
\begin{equation}
\label{eatamsd}
\left<\overline{\delta^2(\Delta)}\right>=\frac{1}{N}\sum_{i=1}^{N}\overline{
\delta_i^2(\Delta)}
\end{equation}
for CTRWs scales like $\langle\overline{\delta^2(\Delta)}\rangle\simeq2D_{\beta}
\Delta/T^{1-\beta}$ \cite{lubelski,he,pccp,neusius}. This linear dependence on
$\Delta$ is preserved for aging CTRWs \cite{johannes}. Apart from this linear
$\Delta$-dependence we also observe the dependence on the process time $T$, a
signature of aging \cite{pt}.

Interestingly, the HDP with diffusivity (\ref{eq-Dx}) displays weak ergodicity
breaking of the form \cite{hdp,fulinski,spain}
\begin{equation}
\left<\overline{\delta^2(\Delta)}\right>=\left(\frac{\Delta}{T}\right)^{1-p}
\left<x^2(\Delta)\right>=\frac{\Gamma(p+1/2)}{\sqrt{\pi}T^{1-p}}\left(\frac{2}{p}
\right)^{2p}\Delta.
\label{eq-tamsd-scaling}
\end{equation}
This behavior is analogous to that of scale-free CTRW motion, despite of the fact
that the increment correlation function of HDPs are (anti)persistent in analogy
to the ergodic FBM \cite{hdp,deng}. We note that also other processes such as
aging and correlated CTRWs \cite{vincent,lomholt} as well as scaled Brownian
motion with time-dependent diffusivity \cite{novikov,sbm-eb} exhibit the duality
between (\ref{msd}) and a linear $\Delta$-dependence of $\overline{\delta^2}$.
Eq.~(\ref{eq-tamsd-scaling}) shows the $T^{p-1}$-scaling as function of the
process time $T$. For subdiffusion with $0<p<1$, that is, the effective
diffusivity of the process decays over time, as the particle ventures into
low-diffusivity areas. Conversely, for $p>1$ the diffusivity increases over time,
the particle discovers areas with increasingly higher $D(x)$.

In the current paper we perform a detailed analysis of the MSD and the time
averaged MSD in the entire range of $\alpha$, including the critical value
$\alpha=2$. We are particularly interested in the extent of the weakly non-ergodic
behavior, especially the fluctuations of the time averaged MSD around
the mean value $\left<\overline{\delta^2}\right>$ characterized by the ergodicity
breaking parameter. Moreover we analyze the non-Gaussianity of the process. In our
analysis we study effects of the scaling exponent $\alpha$ of $D(x)$ as well as
the initial position of the particle. The latter is known to affect the time-scales
at which anomalous diffusion becomes significant \cite{hdp,kehr-87}. We combine
analytical and numerical approaches. The simulations scheme for HDPs was
introduced in Ref.~\cite{hdp}.

\begin{figure*}
\includegraphics[width=18cm]{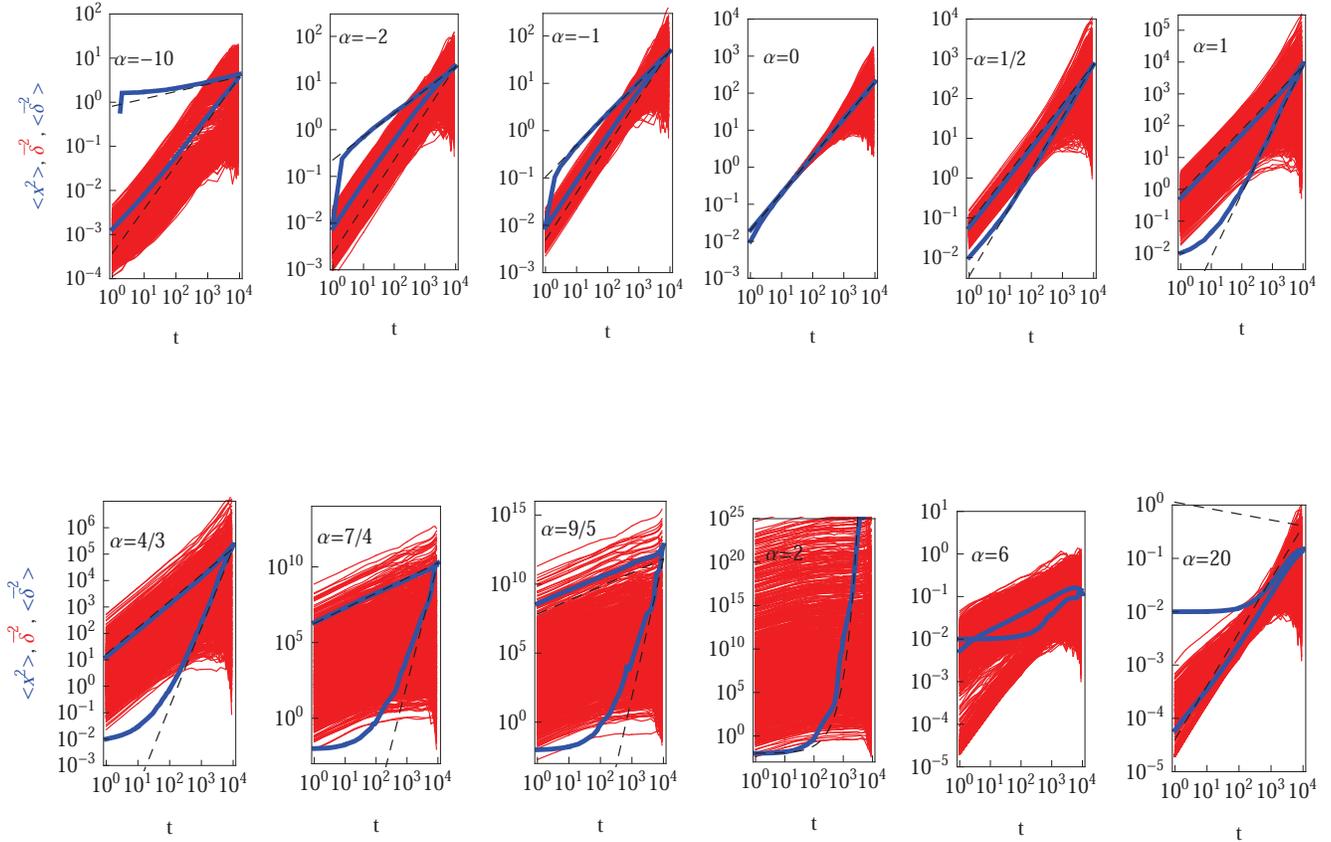}
\caption{MSD of the heterogeneous diffusion process with power-law diffusivity
(\ref{eq-Dx}). The MSD $\langle x^2(t)\rangle$ is represented by the thick blue
curves, whose scaling exponent (\ref{pexp}) varies with the power-law exponent
$\alpha$ of $D(x)$. The individual time averaged MSD traces $\overline{\delta^2}$
appear as thin red curves, and the mean time averaged MSD $\left<\overline{\delta
^2}\right>$ as the thick blue curves, all of which have unit slope, apart from the
critical case $\alpha=2$, where the functional dependence is exponential.
The theoretical asymptotes (\ref{eq-msd-scaling}) for $\langle x^2(t)\rangle$
and (\ref{eq-tamsd-scaling}) for $\left<\overline{\delta^2}\right>$ correspond to
the dashed black lines. The analytical values for $\langle x(\Delta)^2\rangle$ and
$\left<\overline{\delta(\Delta)^2}\right>$ coincide in the limit $\Delta=T$. For
$\alpha=6$ or $p=-1/2$ the theoretical asymptote (\ref{eq-msd-scaling}) does not
hold. We used the following parameters: for each $\alpha$ we show $N=10^3$ traces
of length $T=10^4$, the offset $x_{\mathrm{off}}=0.001$, $D_0=0.01$, and the
starting point $x_0=0.1$.}
\label{fig-msd-tamsd-alfa}
\end{figure*}

\section{Results}
\label{sec-main-results}

In this Section we start with the analysis of the MSD and the time averaged MSD
with its amplitude fluctuations, the latter being quantified by the corresponding
scatter distribution. We then analyze the ergodicity breaking and non-Gaussianity
parameters. Finally, we study the PDF of the function that quantifies the degree
of particle dispersion.

\subsection{MSD and time averages MSD}

Fig.~\ref{fig-msd-tamsd-alfa} shows the results from computer simulations for
the MSD $\langle x^2(t)$ and the time averaged MSD $\overline{\delta^2}$ from
individual realizations along with the average $\left<\overline{\delta^2}
\right>$ taken over all single time traces. The values for the scaling exponent
$\alpha$ of the diffusivity (\ref{eq-Dx}) studied in Fig.~\ref{fig-msd-tamsd-alfa}
cover both the sub- and superdiffusive domains and include, in particular, the
critical value $\alpha=2$ where the scaling exponent $p$ of the MSD (\ref{msd})
diverges. For each $\alpha$ we show $N=10^3$ single trajectories. In all cases,
apart from the critical point, the scaling of the MSD $\langle x^2(t)\rangle$ and
both
individual ($\overline{\delta^2}$) and mean ($\left<\overline{\delta^2}\right>$)
time averaged MSDs agree well with the expected analytical behavior shown by the
dashed lines: the scaling exponent of the MSD varies consistently with $\alpha$,
while that of the time averaged MSD remains unity throughout. The initial
discrepancy between theory and MSDs is due to the choice for the initial position
$x_0=0.1$, whose influence relaxes on a time scale depending on both $x_0$ and
$\alpha$. The deviations of individual time traces $\overline{\delta^2}$ at long
lag times from the predicted behavior is due to unavoidable, bad statistics when
the lag time gets close to the overall length $T$ of the time series.

Irreproducibility of time averages of physical observables such as the MSD is an
intrinsic property of weakly non-ergodic processes \cite{pt,he,pccp,lubelski}.
For CTRW processes with scale-free waiting time distribution individual traces
$x(t)$ contain different, few dominating waiting time events that cause the
amplitude scatter between different trajectories or, in other words, fluctuations
of the apparent effective diffusion constant. This phenomenon occurs no matter
how long the measurement time $T$ is taken. For HDPs the scatter is due to the
difference in the extent of excursions to regions of significantly different
diffusivity.

The amplitude scatter between individual realizations $\overline{\delta^2}$ in our
HDP shown in Fig.~\ref{fig-msd-tamsd-alfa} varies significantly with the value of
the scaling exponent $\alpha$: away from the Brownian value $\alpha=0$ in both
subdiffusive ($\alpha<0$) and superdiffusive ($\alpha>0$) cases the fluctuations
of $\overline{\delta^2}$ become more pronounced when the modulus of $\alpha$
increases. At the critical point the fluctuations of $\overline{\delta^2}$ appear
to diverge, while beyond this critical point, the fluctuations decrease again.
Moreover, a population splitting in a faster (steeper slope of $\overline{\delta
^2}$) and slower (shallower slope) fraction of trajectories appears \cite{hdp},
see especially the panel for $\alpha=6$. In terms of the dimensionless variable
\begin{equation}
\xi=\frac{\overline{\delta^2(\Delta)}}{\left<\overline{\delta^2(\Delta)}\right>}
\end{equation}
the amplitude scatter distribution $\phi(\xi)$ reflects the randomness of
individual time averages of the MSD. For a sub- and a superdiffusive $\alpha$ it
was analyzed in Ref.~\cite{hdp}. As shown here for a whole spectrum of $\alpha$
values, there is a clear trend towards extreme fluctuations at the critical point
$\alpha=2$, but even for considerably smaller values such as $\alpha=7/4$ the
fluctuations around the mean $\langle\overline{\delta^2}\rangle$ are significant.
The width of the fluctuations of $\overline{\delta^2}$ in each panel varies only
marginally with the lag time $\Delta$, apart from the behavior at $\Delta\to T$.
As studied in Ref.~\cite{hdp}, the relative amplitude scatter distribution
$\phi(\xi)$ can be fitted with the generalized Gamma-distribution. In particular,
it tends to zero at $\xi=0$, in contrast to subdiffusive CTRW processes, for which
$\phi(0)$ is always positive, indicating completely stalled trajectories
\cite{he,jae_jpa,pt,pccp}.

We note that when the exponent $\alpha$ approaches the critical value $\alpha=2$,
the number of steps $t^\star$ necessary to approach the theoretically predicted
asymptote (\ref{eq-msd-scaling}) increases significantly. Thus, as seen from
Fig.~\ref{fig-msd-tamsd-alfa}, for $x_0=0.1$ only few simulation steps, $t^\star
\approx2$ are needed for negative $\alpha$ with larger modulus. It increases to
$t^\star\approx10$ steps for $\alpha=1/2$, $t^\star\approx100$ for $\alpha=1$, and
already $t^\star\approx 1000$ for $\alpha=7/4$.

Once the anomalous diffusion exponent $p$ becomes negative, that is, for values of
$\alpha$ larger than the critical value $\alpha=2$, the MSD (\ref{eq-msd-scaling})
becomes a decreasing function of time. This agrees with our simulations if $x_0$
is chosen sufficiently large to enable the relaxation to the theoretical asymptote.
For instance, for the extreme value $\alpha=20$ and values of the initial position
of $x_0=10$ and above the MSD indeed follows the theoretical prediction $\left<
x(t)^2\right>\simeq t^{-1/9}$ (not shown). In this region $\alpha>2$ the
diffusivity grows fast away from the origin the decreasing MSD corresponds to
the localization of particles in regions of slow diffusivity, see the detailed
discussion in Sec.~\ref{sec-spread} below.

In the limit $\alpha\to2$, due to the huge spread one or few extremely large
amplitudes $\overline{\delta^2}$ of the time averaged MSD may substantially
affect the mean $\left<\overline{\delta^2}\right>$. The MSD in this limit
follows an exponential growth $\left<x^2(t)\right>\simeq\exp(2D_0t)$, as
indicated by the dashed curve in the panel for $\alpha=2$ in
Fig.~\ref{fig-msd-tamsd-alfa}. Such a fast increase of the MSD is consistent
with the divergence of the scaling exponent $p$ as function of $\alpha$ in
Eq.~(\ref{eq-msd-scaling}) as well as with the exponential MSD growth predicted
for a parabolically space-varying diffusivity in Refs.~\cite{lubensky07,fulinski}.
This property can be straightforwardly inferred from the diffusion equation,
\begin{equation}
\frac{\partial P(x,t)}{\partial t}=D_0\frac{\partial}{\partial x}\left[(x^2+x_{
\mathrm{off}}^2)\frac{\partial P(x,t)}{\partial x}\right].
\end{equation}
Multiplying both sides with $x^2$ and integrating over $x$ one arrives at
\cite{lubensky07}
\begin{equation}
\left<x^2(t)\right>=x_0^2+\frac{1}{3}x_{\mathrm{off}}\Big[\exp(6D_0t)-1\Big].
\end{equation}
This also rationalizes the observation that the scatter of $\overline{\delta^2}$
is maximal for $\alpha=2$ as the particles perform extremely far-reaching
excursions relative to other $\alpha$-values.

\subsection{Ergodicity breaking parameters}

As introduced in Refs.~\cite{russians-eb,he} the fluctuations of the time averaged
MSD $\overline{\delta^2}$ can be quantified by their variance, the ergodicity
breaking parameter
\begin{equation}
\text{EB}(\Delta)=\lim_{T/\Delta\to\infty}\frac{\left<(\overline{\delta^2(\Delta)})
^2\right>-\left<\overline{\delta^2(\Delta)}\right>^2}{\left<\overline{\delta^2(
\Delta)}\right>^2}.
\label{eq-eb1}
\end{equation}
When $\mathrm{EB}=0$, it means that the process is perfectly reproducible and all
time averages over sufficiently long trajectories give the same value. This case
corresponds to the sharp scatter distribution $\phi(\xi)=\delta(\xi-1)$ \cite{pt,
he,pccp,jae_jpa}. For the canonical Brownian Motion the ergodicity breaking
parameter reaches the zero value in the form $\mathrm{EB}_\mathrm{BM}(\Delta)=4
\Delta/(3T)$ at finite ratio $\Delta/T$. Weakly non-ergodic processes have a
positive value of $\mathrm{EB}$. An alternative, weaker condition for ergodicity
is when the MSD (\ref{msd}) and the time averaged MSD (\ref{eatamsd}) coincide.
To measure the relative deviations from ergodicity, the $\mathcal{EB}$ parameter
was introduced \cite{levywalks1}
\begin{equation}
\mathcal{EB}(\Delta)=\frac{\left<\overline{\delta^2(\Delta)}\right>}{\left<x^2(
\Delta)\right>}.
\label{eq-eb2}
\end{equation}
For ergodic dynamics its value is unity.

\begin{figure}
\includegraphics[width=8cm]{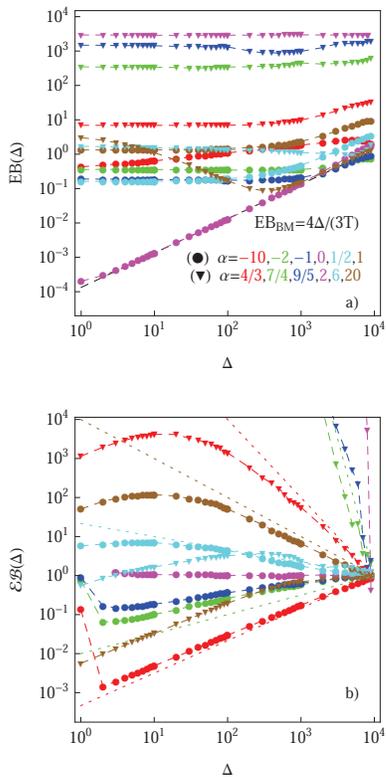}
\caption{Dependence of the ergodicity breaking parameters a) EB and b) $\mathcal{
EB}$ on the lag time $\Delta$. The Brownian asymptote $\mathrm{EB}_\mathrm{BM}(
\Delta)$ is shown as the dashed curve in panel a). The asymptotes (\ref{eq-eb2})
are shown in panel b) as dashed lines of the corresponding color. The initial
position of the particle was $x_0=0.1$ and $N=3\times 10^3$ traces were used for
each value of $\alpha$, all other parameters are the same as those used in
Fig.~\ref{fig-msd-tamsd-alfa}. The symbols in both panels correspond to the
same parameters.}
\label{fig-eb-lag}
\end{figure}

\begin{figure}
\includegraphics[width=7.0cm]{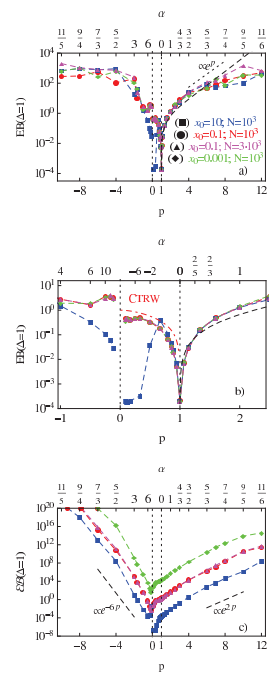}
\caption{Dependence on the scaling exponent $p$, Eq.~(\ref{pexp}), of $\mathrm{
EB}(\Delta=1)$ in panels a) and the magnification in panel b) as well as $\mathcal{
EB}(\Delta=1)$ in panel c). The theoretical predictions from Ref.~\cite{hdp}
correspond to the dashed black curve in panels a) and b). The ergodicity breaking
parameter of the subdiffusive CTRW given by Eq.~(\ref{eq-eb-ctrw}) is shown by the
dot-dashed red curve in panel b). To compute each point in the graphs for $N=10^3$
traces of $T=10^4$ steps takes some two hours on a standard 3 GHz core workstation.
Different sets of the shown points correspond to varying particle initial
conditions $x_0$ indicated by different colors. Note the alternative scale for the
power exponent $\alpha=2-2/p$
on the top axes. Parameters are the same as in Fig.~\ref{fig-msd-tamsd-alfa}.}
\label{fig-eb-p}
\end{figure}

To calculate the $\mathrm{EB}$ parameter analytically is not always an easy task.
To see this, note that the MSD follows from the change of the stochastic variable
to the standard Wiener process in the form $y(x(t))=\int^{x'}dx'[2D(x')]^{-1/2}$
\cite{hdp}. For the time averaged MSD $\overline{\delta^2}$ the calculation is
already more complicated, as it involves the two-point position correlation
function. The latter is expressed via Fox $H$-functions for the HDP \cite{hdp}.
The analytical derivation of the ergodicity breaking parameter $\mathrm{EB}$,
however, requires fourth-order moments, whose calculation is a formidable task.
So far only approximate methods are known for HDPs \cite{hdp}. For subdiffusive
CTRW processes it is possible to obtain the $\mathrm{EB}$ parameter more easily
from the conjectured and numerically proven equivalence $\xi=\overline{\delta^2}/
\left<\overline{\delta^2}\right>\equiv n(t)/\langle n(t)\rangle$ of $\xi$ with the
ratio of the number of steps $n(t)$ in an individual realization and the average
$\langle n(t)\rangle$ \cite{he,johannes}. As for the limiting distribution for
$n$ is known \cite{hughes}, this allows straightforward calculation of $\phi(\xi)$
and its moments \cite{he,johannes}. For the multiplicative process studied here
such a scheme does not work. We also note that another parameter involving the
fourth moment of the particle displacement is the non-Gaussianity parameter $G$
\cite{franosch13} discussed below. Due to the lack of an analytical theory, a
major reason for the current simulations study is to explore the behavior of
these two parameters in the whole range of the model parameters. 

Figs.~\ref{fig-eb-lag} and \ref{fig-eb-p} display the dependence of the ergodicity
breaking parameters on the lag time $\Delta$, the initial position $x_0$, as well
as the scaling exponent $p$. We find that for standard Brownian motion with $\alpha
=0$ the ergodicity breaking parameter EB follows the known asymptote $\mathrm{EB}
(\Delta)=\mathrm{EB}_{\mathrm{BM}}(\Delta)$ from above. Concurrently $\mathcal{EB}
(\Delta)\to1$ as expected for ergodic motion, except for very small $\Delta$ values
because of the initial relaxation of the influence of the initial position $x_0$.
As we depart from the value $\alpha=0$ of Brownian Motion, the magnitude of
$\mathrm{EB}$ grows and its functional dependence on the lag time becomes less
pronounced, see Fig.~\ref{fig-eb-lag}a. Approaching the critical point $\alpha
=2$, due to the huge fluctuations of $\overline{\delta^2}$ and the ensuing
possibility of extreme events the ergodicity breaking parameter also explicitly
depends on the number of traces $N$ used for the averaging and reaches values
of $\mathrm{EB}\sim10^4$ and higher. On both sides of the critical point,
corresponding to large positive or negative values of $p$, the
values of the parameter $\mathrm{EB}(\Delta=1)$ approach one another, as seen in
Fig.~\ref{fig-eb-p}a, while $\mathcal{EB}(\Delta=1)$ exhibits a jump,
Fig.~\ref{fig-eb-p}b.  

The ratio of the different MSDs given by $\mathcal{EB}$ also depends on the power
exponent $\alpha$ and the initial value $x_0$. As anticipated already from
Fig.~\ref{fig-msd-tamsd-alfa}, for the chosen initial condition $x_0=0.1$
the magnitude of $\mathcal{EB}$ decreases as $\alpha$ gets progressively negative,
while $\mathcal{EB}$ grows as $\alpha$ increases from 0 to 2, reaching very large
values $\mathcal{EB}(\Delta/T\to0)$ at $\alpha=2$. This is illustrated in
Fig.~\ref{fig-eb-lag}b which is also consistent with the scaling
\begin{equation}
\mathcal{EB}(\Delta)\simeq\left(\frac{\Delta}{T}\right)^{1-p}
\end{equation}
with the lag time at different $p=2/(2-\alpha)$ values. Because of the initial
condition $x_0$, similar to our statements for $\left<\overline{\delta^2}\right>$
this asymptote is approached later when $\alpha\to2$.     

In Fig.~\ref{fig-eb-p} we also analyze the effects of the initial position $x_0$
and show the dependence of the ergodicity breaking parameters for long traces or
short lag times, i.e., when $\text{EB}(\Delta=1)$ which is practically equivalent
to $\text{EB}(\Delta/T\to0)$, versus the anomalous diffusion exponent $p$ and the
scaling exponent $\alpha$. We observe that in the region $p>1$ the values of
$\mathrm{EB}(\Delta=1)$ obtained from simulations is in good agreement with the
analytical estimate $\text{EB}(\Delta/T\to0)$ from Ref. \cite{hdp}, represented
by the dashed curve in  Fig.~\ref{fig-eb-p}a and b. These approximations are
based on the asymptotic scaling of $D(x)$ and do not consider the effect of the
initial positions $x_0$, assumed to be relaxed in the relevant $\Delta/T\to0$
limit. To see their actual impact on the dynamics we study the ergodicity
breaking parameters numerically.

We observe that the value of $\mathrm{EB}(\Delta=1)$ grows from the small
pre-asymptotic Brownian value at $p=1$ to progressively larger values at larger
modulus of $p$, as $\alpha$ approaches the critical value $\alpha=2$ from below
and above. The functional dependence of $\mathrm{EB}(\Delta=1,p)$ obtained from
our simulations follows quite well the predictions from the approximate calculation
of $\mathrm{EB}$ \cite{hdp}, particularly in the range $0<\alpha\lesssim3/2$, see
Fig.~\ref{fig-eb-p}b. The deviations for even more pronounced variation of
$D(x)$ as $\alpha$ tends to 2 are likely due to insufficient statistics. As we
show in Fig.~\ref{fig-eb-p}a by the circles and triangles, the value $\mathrm{EB}(
\Delta=1,p\gg 1)$ reveals measurable deviations for $N=3\times10^3$ as compared to
$N=10^3$ traces used for the averaging, while for less extreme $\alpha$ values the
two sets yield nearly identical results. This is a further indication towards a
divergence of the fluctuations at the critical point. Namely, the chance to find
an even larger amplitude of $\overline{\delta^2}$ in a given realization grows
with the number of simulated trajectories.

As already anticipated in Ref.~\cite{hdp} the HDP process considered here is in
some aspects reminiscent of subdiffusive CTRW processes with scale-free, power-law
waiting time distributions with regard to the scaling of the MSD and the time
averaged MSD, compare also Sec.~\ref{sec-discussion}. Apparently, the ergodicity
breaking properties of these two processes also reveal similar features: the
fluctuations of $\overline{\delta^2}$ increase when the anomalous diffusion
exponent $p$ becomes successively smaller than the Brownian value 1. Indeed, as
shown
in Fig.~\ref{fig-eb-p}b for the relevant region $0<p<1$ there exists even a close
agreement of the quantitative behavior of CTRW and HDP. We show the ergodicity
breaking parameter obtained for the CTRW process \cite{he}
\begin{equation}
\mathrm{EB}_{\mathrm{CTRW}}(\Delta/T\to0)=\frac{2\Gamma(1+\beta)^2}{\Gamma(1+2
\beta)}-1,
\label{eq-eb-ctrw}
\end{equation}
where $0<\beta<1$ is the exponent in the PDF $\psi(\tau)\simeq\tau^{-1-\beta}$
of the waiting times $\tau$ \cite{report}. For the CTRW process the exponent
$\beta$ also occurs in the MSD (\ref{msd}). We note that in contrast to HPDs in
superdiffusive CTRW processes the nonergodic behavior is merely ultraweak,
effecting a different prefactor in the time averaged MSD compared to the MSD
\cite{levywalks,levywalks1}.

For $\alpha\lesssim-2$ we find that $\mathrm{EB}(\Delta=1)\sim0.4$ for initial
positions $x_0$ close to the origin, in agreement with the results presented
in Ref.~\cite{hdp}. For $\alpha>2$ or $p<0$ the analytical model of Ref.~\cite{hdp}
no longer applies but the simulations yield another region of growth for the value
of $\mathrm{EB}(\Delta=1)$. Note that $\mathrm{EB}(\Delta=1)$ also depends on the
trace length $T$ (not shown), as discussed for two-dimensional HDPs \cite{hdp}.
As we show in Fig.~\ref{fig-eb-p}c, the magnitude of $\mathcal{EB}(\Delta=1)$
strongly varies with $x_0$. In the region $p\gg1$ the ratio of the MSDs grows as
$\mathcal{EB}(\Delta=1,p)\propto\exp(2p)$ indicated by the dashed asymptote in
Fig.~\ref{fig-eb-p}c, while for large negative $p$ we observe $\mathcal{EB}(
\Delta=1,p)\propto\exp(-6p)$. We finally note that $\mathcal{EB}(\Delta=1)$ reveals
a much weaker dependence on the number $N$ of simulated traces, see the circles and
triangles in Fig.~\ref{fig-eb-p}c. The reason is that $\mathcal{EB}$ involves
only the second moment of the time averaged MSD, while the $\mathrm{EB}$ is defined
in terms of the fourth order moment, which is more sensitive to large variations.

\subsection{Non-Gaussianity parameter}

\begin{figure}
\includegraphics[width=8.4cm]{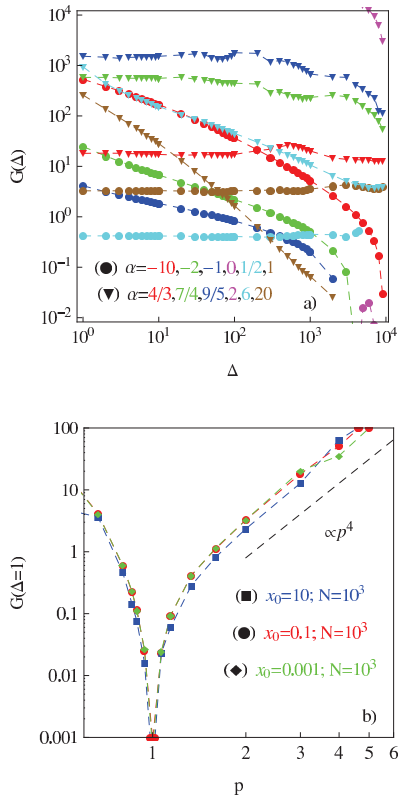}
\caption{\textbf{a)} Non-Gaussianity parameter $G(\Delta)$ as function of the
lag time, plotted for the parameters of Fig.~\ref{fig-eb-lag} with $N=10^3$
trajectories for each $\alpha$ value. The approximately zero-valued trace
$G(\Delta)$ for Brownian motion ($\alpha=0$) is only shown partially. Color
coding is the same as in Fig.~\ref{fig-eb-lag}. \textbf{b)} The value of
$G(\Delta=1)$ evaluated for varying power exponent $\alpha$ and initial
position $x_0$. The color scheme is the same as in Fig.~\ref{fig-eb-p}.}
\label{fig-non-gauss}
\end{figure}

The non-Gaussianity parameter $G$ is a sensitive experimental indicator that often
enables one to distinguish the type of particular diffusion processes observed
in single-particle tracking experiments \cite{weiss14-pccp}. It is related to the
stationarity of increments of the diffusion process and involves the fourth moment
along the time-averaged trajectory. For the diffusion process in an embedding space
of dimension $d$ it is defined via the experimentally relevant time averaged
quantities as \cite{franosch13}
\begin{equation}
G(\Delta)=\frac{d}{d+2}\times\frac{\left<\overline{\delta^4(\Delta)}\right>}{\left<
\overline{\delta^2(\Delta)}\right>^2}-1,
\end{equation}
where, in analogy to Eq.~(\ref{eq-tamsd}), the fourth moment is defined via
\begin{equation}
\overline{\delta^4(\Delta)}=\frac{1}{T-\Delta}\int_0^{T-\Delta}\Big[x(t+\Delta)
-x(t)\Big]^4~dt.
\end{equation}
For Brownian and fractional Brownian motion, both Gaussian processes, one
consistently finds $G=0$. For diffusion processes revealing a transient anomalous
diffusion behavior, this parameter deviates substantially from zero, see, e.g.,
the discussion in Ref.~\cite{weiss14-pccp}.

\begin{figure*}
\includegraphics[width=18cm]{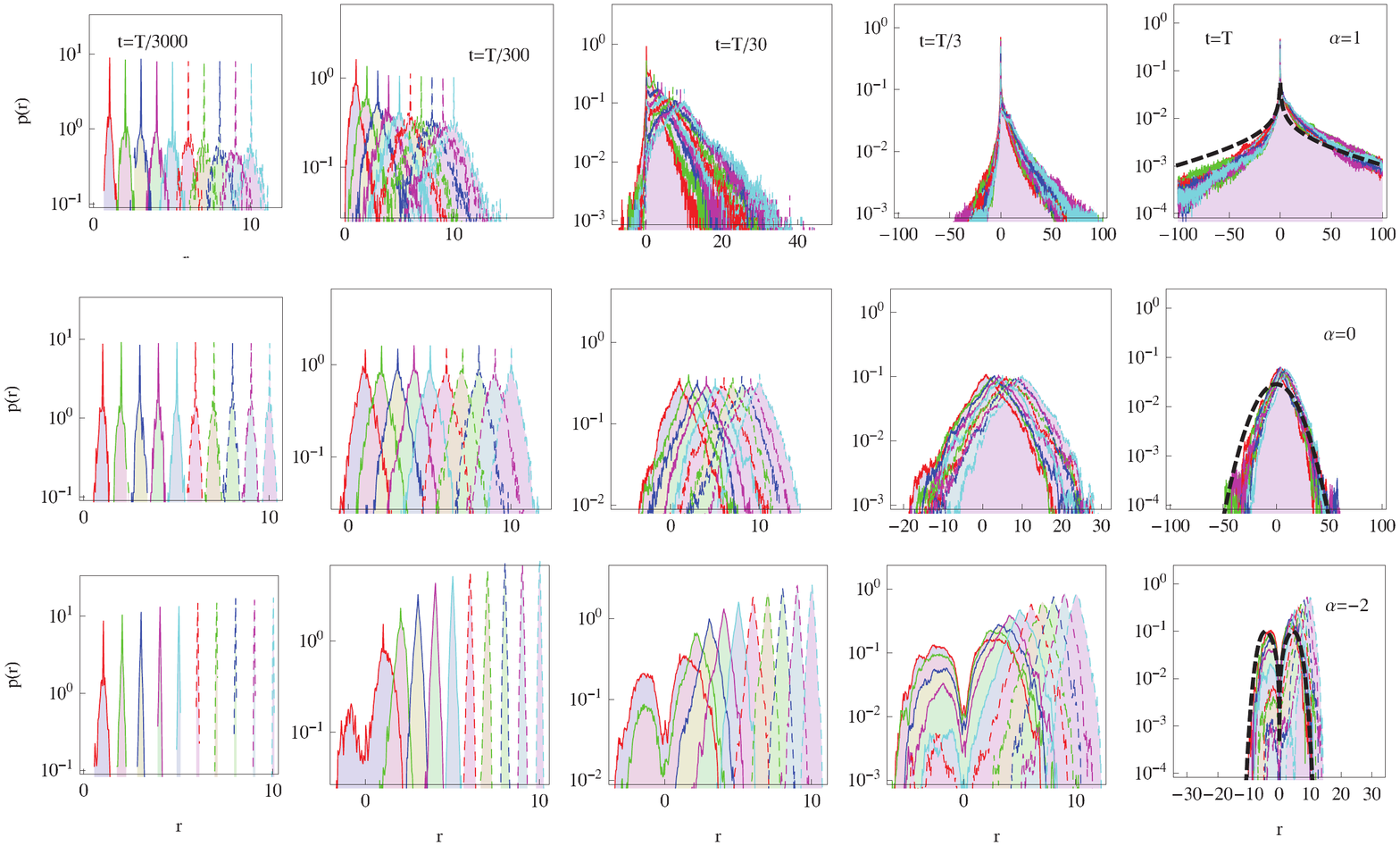}
\caption{Spreading of diffusing particles for ten equidistantly placed initial
positions. The dashed curves are the theoretical PDFs given by Eq.~(\ref{PDF-HDP})
in the diffusion limit. Parameters: $T=10^4$, averaged over $N=200$ trajectories
for each initial position $x_0$, and $5000$ bins were used in the range $-100<x
<100$.}
\label{fig-spread}
\end{figure*}

We systematically examine the behavior of $G(\Delta)$ in Fig.~\ref{fig-non-gauss}a.
Corroborating the results for the ergodicity breaking parameter, for Brownian
motion we obtain approximately zero values. As the power $\alpha$ in $D(x)\sim|x|^{
\alpha}$ deviates from zero, the non-Gaussianity parameter reveals a rich behavior
as a function of the lag time $\Delta$ and the initial position $x_0$.

More specifically, in the region $\alpha<0$ the non-Gaussianity parameter
progressively grows and reaches considerably large values for large $|\alpha|$,
see the curve for $\alpha=-10$ in Fig.~\ref{fig-non-gauss}a. The value $G(\Delta)$
systematically decreases with the lag time $\Delta$ along the trace for negative
$\alpha$. In the region $0<\alpha<2$ the value of $G$ also grows with $\alpha$ but,
in contrast to the case $\alpha<0$, the function $G(\Delta)$ stays rather constant
with $\Delta$. As we approach the critical $\alpha=2$ the non-Gaussianity parameter
reaches very high values. In the region $\alpha>2$ the value of $G(\Delta)$
decreases again. These features of the functional behavior of $G(\alpha=2-2/p)$
correspond with the properties of $\mathrm{EB}(\alpha)$, compare the curves in
Fig.~\ref{fig-non-gauss}a and \ref{fig-eb-lag}a. Note that similar to $\mathrm{EB}$
the statistics required to obtain a smooth curve for $G(\Delta)$ strongly depends
on $\alpha$: the result from $N=10^3$ traces in Fig.~\ref{fig-non-gauss}a acquires
pronouncedly higher fluctuations when we approach the critical value $\alpha=2$. 

We find that variations in the initial positions $x_0$ only have a marginal effect
on $G(\Delta=1,p)$. In Fig.~\ref{fig-non-gauss}b we illustrate the range of $G$
values for exponents in the range $-4<\alpha<5/3$. Even for the wide range from
$x_0=10$ to $x_0=0.001$ the value of $G(\Delta=1)$ does not reveal any appreciable
variation, in stark contrast to the strong $x_0$-dependence of $\mathrm{EB}(\Delta
=1)$. This indicates that for the HDP the non-Gaussianity parameter is more
robust than $\mathrm{EB}$ with respect to the choice of the initial condition. In
Fig.~\ref{fig-non-gauss}b the dashed line indicates the proportionality to $p^4$,
which nicely matches the measured shape of $G(\Delta=1,p)$.

\subsection{Probability density function and particle focusing}
\label{sec-spread}

We finally analyze the time evolution of the PDF of the particle position in
Fig.~\ref{fig-spread} for three different values of the scaling exponent $\alpha$,
distinguishing superdiffusion (top row for $\alpha=1$, i.e., ballistic motion
with $p=2$), normal diffusion (middle row for $\alpha=0$ and $p=1$), and
subdiffusion (bottom row for $\alpha=-4$, i.e., $p=1/3$). In each row the leftmost
panel represents a very early evolution of the PDF close to the initial condition,
while in the rightmost panel the PDF in some of the cases has almost reached the
diffusion limit, in which the analytical asymptote (\ref{PDF-HDP}) is valid.

We observe that for HDPs with positive $\alpha$, i.e., when the diffusivity $D(x)$
grows with the modulus of the position, the PDF for particles with an initial
position far away from the origin, an asymmetric shape of the PDF is effected.
Namely, they progressively move towards regions of small diffusivity and accumulate
there. This focusing due to the quenched nature of the diffusivity erases any
memory of the initial condition and the common asymptote (\ref{PDF-HDP}) of
stretched Gaussian shape ($0<\alpha<2$) is approached. In the opposite case with
negative $\alpha$ the focusing of particles in lower diffusivity regions applies
again, albeit at longer times than shown here. This time, however, instead of the
cusp at the origin for positive $\alpha$, the PDF drops down to zero at the origin
and acquires the bimodal shape predicted by (\ref{PDF-HDP}), a compressed Gaussian.
For the Brownian case with vanishing $\alpha$, no focusing takes place. In the
final panel the distribution is already so broad that the individual, shifted PDFs
appear on top of each other.

Similarly to the evolution of the MSD shown in Fig.~\ref{fig-msd-tamsd-alfa},
the relaxation time required for the system of diffusing particles to approach
this long-time limit, $P(x,t)$ depends on the power $\alpha$ of the diffusivity
in Eq.~(\ref{eq-Dx}). For more negative $\alpha$ values the traces are not yet
relaxed to the theoretical long-time shape even after $T=10^4$ steps. The
particles starting at larger $x_0$ remain trapped for the whole length of the
simulation. Conversely, when the diffusivity increases fast away from the origin
(moderate positive $\alpha$ values), the equilibration to the long-time PDF is
relatively fast, see the panel for $\alpha=1$ in Fig.~\ref{fig-spread}.

\section{Discussion and Conclusions}
\label{sec-discussion}

For HPD processes whose diffusivity $D(x)$ varies in power-law form with the
distance $x$ from the origin and which give rise to anomalous diffusion we
analyzed in detail the weakly non-ergodic behavior for varying power exponents
$\alpha$ and initial positions $x_0$ of the particles. In particular, we examined
the functional dependencies and magnitudes of the variation of the ensemble and
time averaged characteristics of such HDPs with these parameters. The fluctuations
of the amplitude of the time averaged MSD of individual time traces were shown to
systematically grow with increasing departure of the anomalous diffusion exponent
$p$ from the Brownian value $p=1$, corresponding to the scaling exponent $\alpha=0$
in the power-law form
for $D(x)$. The fluctuations increase dramatically when the scaling exponent
approaches its critical value $\alpha=2$. At that critical point we corroborated
the turnover from the power-law scaling in time of the MSD to an exponential
growth. Beyond the critical point the fluctuations of the time averaged MSD become
smaller again. Concurrently we observe a distinct population splitting into a
faster and slower fraction of time averaged MSDs.

We paid particular emphasis on several parameters used to classify the departure
from ergodic behavior, namely the two ergodicity breaking parameters $\mathrm{EB}$
and $\mathcal{EB}$ as well as the non-Gaussianity parameter $G$. While $\mathcal{
EB}$ is simply defined as the ratio of the mean time averaged MSD versus the MSD,
both $\mathrm{EB}$ and $G$ are based on the fourth order moments of the particle
position and are known analytically only from approximate theories. A detailed
numerical analysis of their properties was therefore used to obtain more concrete
information on their behavior for different values of the scaling exponent $\alpha$
and the initial particle position $x_0$ in the heterogeneous environment
\cite{REM}. Within
the analyzed parameter range we find that the behavior of $\text{EB}(\Delta=1,x_0)$
is indeed in agreement with the heuristic theoretical analysis from Ref.~\cite{hdp}.
The parameter portraits for the ergodicity breaking and non-Gaussianity parameters
obtained from our numerical analysis will be useful for actual data evaluation. We
also explored the detailed behavior of the HDP dynamics at the critical value
$\alpha=2$ and its vicinity, in particular with respect to the dramatic values
reached by $\mathrm{EB}$ reflecting the dramatic fluctuations of the time averaged
MSD. A systematic numerical analysis of the particle PDF for varying initial
conditions $x_0$ sheds additional light on HDPs with different exponent $\alpha$,
in particular, the visualization of the particle focusing in low diffusivity
regions.

In HDPs the non-ergodic behavior arises due to the heterogeneity of the
environment. Physically, this represents a space-dependent mobility or
temperature. It is not a property of the particle, and each time the particle
revisits a given point $x$ in space it has the same diffusivity $D(x)$. In a
random walk sense, this scenario could also be translated into a local
dependence of the waiting time for a jump event. In that interpretation the HDP
corresponds to a motion in a quenched energy landscape \cite{bouchaud}, albeit
a deterministic (in contrast to random) one. As such, the process is intrinsically
different from renewal CTRW processes, which correspond to the motion in an
annealed environment \cite{bouchaud}.

Nevertheless we observed that subdiffusive HDPs share a number of features
with subdiffusive CTRWs with scale-free, power-law waiting time distribution.
These features include the scaling laws for the ensemble and time averaged MSDs,
and the form of the ergodicity breaking parameter $\mathcal{EB}$ (in the relevant
range $0<p<1$). From the sole analysis of the MSDs $\langle x^2(t)\rangle$ and
$\overline{\delta^2}$ as well as the ergodicity breaking parameters, a significant
distinction between CTRW and HDP is therefore difficult. Yet there exist some
crucial differences between the HDP and CTRW processes, that can be measured.
Thus, in HDPs with $0<p<1$ the distribution $\phi(\xi)$ of the relative amplitude
$\overline{\delta^2}$ of individual realizations decays to zero at $\xi=0$,
in contrast to the CTRW's finite fraction of immobile particles reflected in the
finite value $\phi(0)>0$. Indeed the scatter distribution $\phi(\xi)$
was previously advocated as a good diagnosis tool for different stochastic
processes \cite{pt,jae_jpa}, complementing other stochastic analysis methods
\cite{methods,weiss14-pccp,thiel}.

HDPs are physically meaningful alternatives to the standard stochastic models for
anomalous diffusion processes, namely, the CTRW process with long-tailed waiting
time distribution, FBM and fractional Langevin equation motion based on Gaussian
yet long-ranged in time correlations, diffusion in fractal environments, or their
combinations \cite{granules,insulin,channels,thiel}.
HPDs are weakly non-ergodic, sharing some yet not all properties of CTRW processes. 
It is therefore important to explore the properties of HDPs in more detail in the
future as well as to develop more sophisticated tools to interpret measured time
series and with confidence identify the underlying stochastic mechanism. In
particular, we will study the properties of confined and aged HDPs as well as
spatially varying diffusivities with a random component, and the coupling of HDPs
with active motion.

\acknowledgments

The authors acknowledge funding from the Academy of Finland (FiDiPro scheme
to RM) and the German Research Council (DFG Grant CH 707/5-1 to AGC).

\end{document}